\documentclass[amsmath,amssymb,superscriptaddress,nobalancelastpage,aps,prl,twocolumn]{revtex4-2}
\usepackage{varioref}
\usepackage{xr-hyper}
\usepackage{xcolor}
\usepackage{nicefrac}
\usepackage{xfrac}
\usepackage{hyperref}
\usepackage{lineno}
\usepackage{graphicx}
\usepackage{amssymb}
\usepackage{natbib}
\usepackage{calc}
\usepackage{SIunits}
\usepackage{textcomp}
\usepackage{amsmath}
\usepackage{amssymb}
\usepackage{color}
\usepackage{ulem}
\usepackage{bm}
\hypersetup{colorlinks,linkcolor=blue,urlcolor=blue,citecolor=blue}

\begin{document}

\title{Spin Fluctuations in Sr$_{1.8}$La$_{0.2}$RuO$_4$}

\author{Zheng~He}
\affiliation{
State Key Laboratory of Surface Physics and Department of Physics, Fudan University, Shanghai 200433, China
}

\author{Qisi~Wang}
\email{qwang@cuhk.edu.hk}
\affiliation{
	State Key Laboratory of Surface Physics and Department of Physics, Fudan University, Shanghai 200433, China
}
\affiliation{
	Department of Physics, The Chinese University of Hong Kong, Shatin, Hong Kong, China
}

\author{Yu~Feng}
\affiliation{
State Key Laboratory of Surface Physics and Department of Physics, Fudan University, Shanghai 200433, China
}
\affiliation{
Institute of High Energy Physics, Chinese Academy of Sciences (CAS), Beijing 100049, China
}
\affiliation{
Spallation Neutron Source Science Center (SNSSC), Dongguan 523803, China
}

\author{Chul Kim}
\affiliation{
Institute of Physics and Applied Physics, Yonsei University, Seoul, 03722, Korea
}

\author{Wonshik Kyung}
\affiliation{
Center for Correlated Electron Systems, Institute for Basic Science, Seoul 08826, Korea
}
\affiliation{
Department of Physics and Astronomy, Seoul National University, Seoul 08826, Korea
}

\author{Changyoung Kim}
\affiliation{
Center for Correlated Electron Systems, Institute for Basic Science, Seoul 08826, Korea
}
\affiliation{
	Department of Physics and Astronomy, Seoul National University, Seoul 08826, Korea
}

\author{Hongliang~Wo}
\affiliation{
	State Key Laboratory of Surface Physics and Department of Physics, Fudan University, Shanghai 200433, China
}
\affiliation{
	Shanghai Qi Zhi Institute, Shanghai 200232, China
}

\author{Gaofeng~Ding}
\affiliation{
	State Key Laboratory of Surface Physics and Department of Physics, Fudan University, Shanghai 200433, China
}
\affiliation{
	Shanghai Qi Zhi Institute, Shanghai 200232, China
}

\author{Yiqing~Hao}
\affiliation{
	State Key Laboratory of Surface Physics and Department of Physics, Fudan University, Shanghai 200433, China
}

\author{Feiyang~Liu}
\affiliation{
	State Key Laboratory of Surface Physics and Department of Physics, Fudan University, Shanghai 200433, China
}

\author{Helen~C.~Walker}
\affiliation{
ISIS Facility, STFC Rutherford Appleton Laboratory, Chilton, Didcot, Oxfordshire OX11 0QX, United Kingdom
}

\author{Devashibhai~T.~Adroja}
\affiliation{
ISIS Facility, STFC Rutherford Appleton Laboratory, Chilton, Didcot, Oxfordshire OX11 0QX, United Kingdom
}
\affiliation{
Highly Correlated Matter Research Group, Physics Department, University of Johannesburg, P.O. Box 524, Auckland Park 2006, South Africa
}

\author{Astrid~Schneidewind}
\affiliation{
J\"{u}lich Centre for Neutron Science (JCNS) at Heinz Maier-Leibnitz Zentrum (MLZ), Forschungszentrum J\"{u}lich GmbH, Lichtenbergstr.$_{ }$1, 85748 Garching, Germany
}

\author{Wenbin~Wang}
\affiliation{
	Institute for Nanoelectronic Devices and Quantum Computing, Fudan University, Shanghai 200433, China
}

\author{Jun~Zhao}
\email{zhaoj@fudan.edu.cn}
\affiliation{
State Key Laboratory of Surface Physics and Department of Physics, Fudan University, Shanghai 200433, China
}
\affiliation{
Institute for Nanoelectronic Devices and Quantum Computing, Fudan University, Shanghai 200433, China
}
\affiliation{
Shanghai Qi Zhi Institute, Shanghai 200232, China
}
\affiliation{
Shanghai Research Center for Quantum Sciences, Shanghai 201315, China
}


\begin{abstract}
We use inelastic neutron scattering to study spin fluctuations in Sr$_{1.8}$La$_{0.2}$RuO$_4$, where Lanthanum doping triggers a Lifshitz transition by pushing the van Hove singularity in the $\gamma$ band to the Fermi energy.
Strong spin fluctuations emerge at an incommensurate wave vector $\mathbf{Q}_{ic} = (0.3,0.3)$, corresponding to the nesting vector between $\alpha$ and $\beta$ Fermi sheets. The incommensurate antiferromagnetic fluctuations shift toward $(0.25,0.25)$ with increasing energy up to ${\sim}110$~meV.
By contrast, scatterings near the ferromagnetic wave vectors $\mathbf{Q} = (1,0)$ and $(1,1)$ remain featureless at all energies. This contradicts the weak-coupling perspective that suggests a sharp enhancement of ferromagnetic susceptibility due to the divergence of density of states in the associated $\gamma$ band.
Our findings imply that ferromagnetic fluctuations in Sr$_2$RuO$_4$ and related materials do not fit into the weak-coupling paradigm, but instead are quasi-local fluctuations induced by Hund's coupling.  This imposes significant constraints for the pairing mechanism involving spin fluctuations.
\end{abstract}

\maketitle

The pairing mechanism of the Sr$_2$RuO$_4$ superconductor has been the focus of tremendous research activities~\cite{MackenzieRMP2003,MaenoJPSJ2012,KallinRPP2016,MackenzieNPJQM2017}, but as yet remains a mystery.
For a long time, Sr$_2$RuO$_4$ has stood as a promising candidate for a spin-triplet superconductor with a chiral $p$-wave order parameter~\cite{RiceJPCM1995,BaskaranPBCM1996,RaghuPRL2010}.  A series of experiments supported this belief.
Early on, nuclear magnetic resonance (NMR)~\cite{IshidaNat1998} and polarized neutron diffraction~\cite{DuffyPRL2000} measurements show an unchanged magnetic susceptibility across the superconducting transition temperature ($T_c$), suggesting an odd pairing state.
In addition, a spontaneous time-reversal symmetry breaking at $T_c$ has been observed by muon spin relaxation~\cite{LukeNat1998} and polar Kerr effect~\cite{XiaPRL2006} studies, which reveals a chiral character of the superconducting state.

\begin{figure}[b]
\centering
\includegraphics[width=0.45\textwidth]{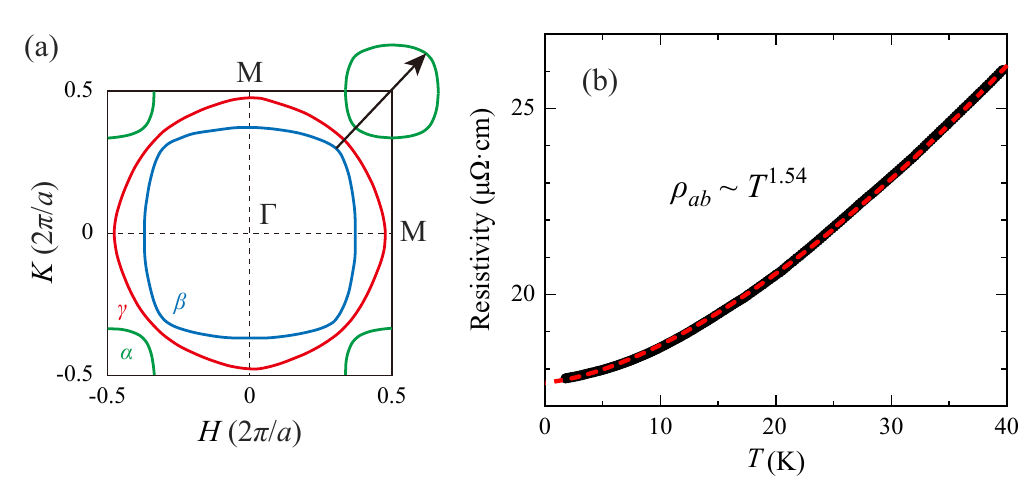}
\caption{
(a) Schematic of the Fermi surfaces in the first Brillouin zone of Sr$_{2-y}$La$_y$RuO$_4$ close to the Lifshitz transition, inferred from ARPES measurements~\cite{ShenPRL2007}. Green, blue and red solid curves represent $\alpha$, $\beta$, and $\gamma$ Fermi sheets, respectively. Black arrow indicates the nesting vector between $\alpha$ and $\beta$ bands.
(b) In-plane resistivity $\rho_{ab}$ versus temperature $T$ in Sr$_{1.8}$La$_{0.2}$RuO$_4$. Red dashed curve is a power-law fitting $AT^n$ + $\rho_{0}$ with $n = 1.54 \pm 0.04$.
}
\label{fig1}
\end{figure}

\begin{figure*}
\centering
\includegraphics[width=0.75\textwidth]{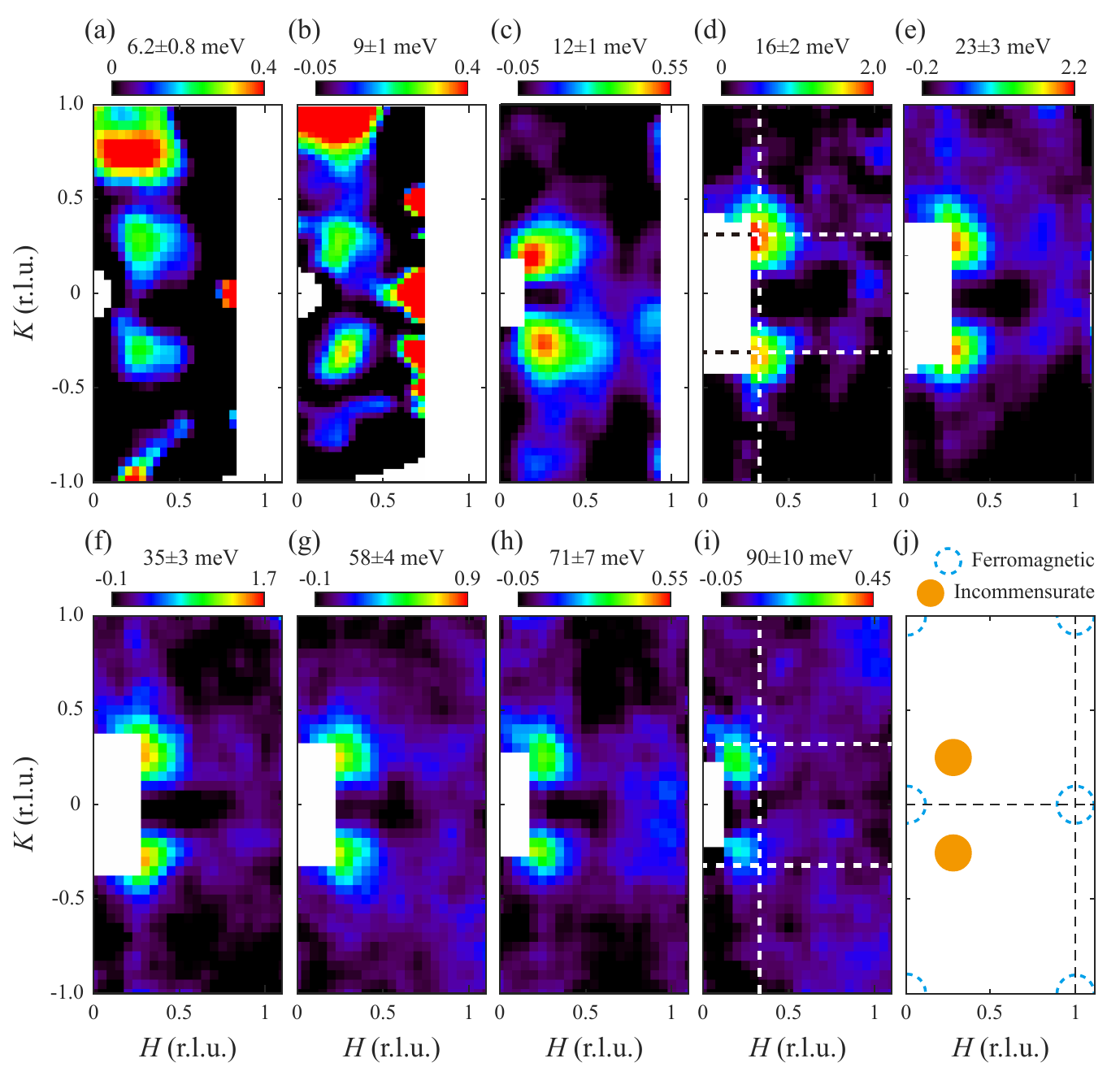}
\caption{
Momentum dependence of the spin fluctuations in Sr$_{1.8}$La$_{0.2}$RuO$_4$ at 5~K. Constant-energy images at indicated energies measured with incident neutron energies of $E_i$ = 18.8~(a) and (b), 40.2~(c), and 135.8~meV (d)–(i). Contour maps in (a) and (b) are rotated clockwise by $90^\circ$ according to the $C_4$ crystal symmetry for direct comparisons with data collected with higher incident energies. Data are symmetrized with respect to the $K$ axis to enhance statistical accuracy and the $|\mathbf{Q}|$-dependent background is subtracted following the method introduced in Ref.~\cite{WangNC2016}. Color bars indicate intensity in unit of mbarn sr$^{-1}$ meV$^{-1}$ f.u.$^{-1}$. Dashed lines in (d) and (i) mark the momenta with $H=0.3$ and $K=\pm0.3$. The antiferromagnetic signals at 90 meV locate clearly at smaller wave vectors. (j) Schematic representation of the incommensurate antiferromagnetic and ferromagnetic wave vectors in the $(H,K)$ plane.
}
\label{fig2}
\end{figure*}

\begin{figure}
\centering
\includegraphics[width=0.45\textwidth]{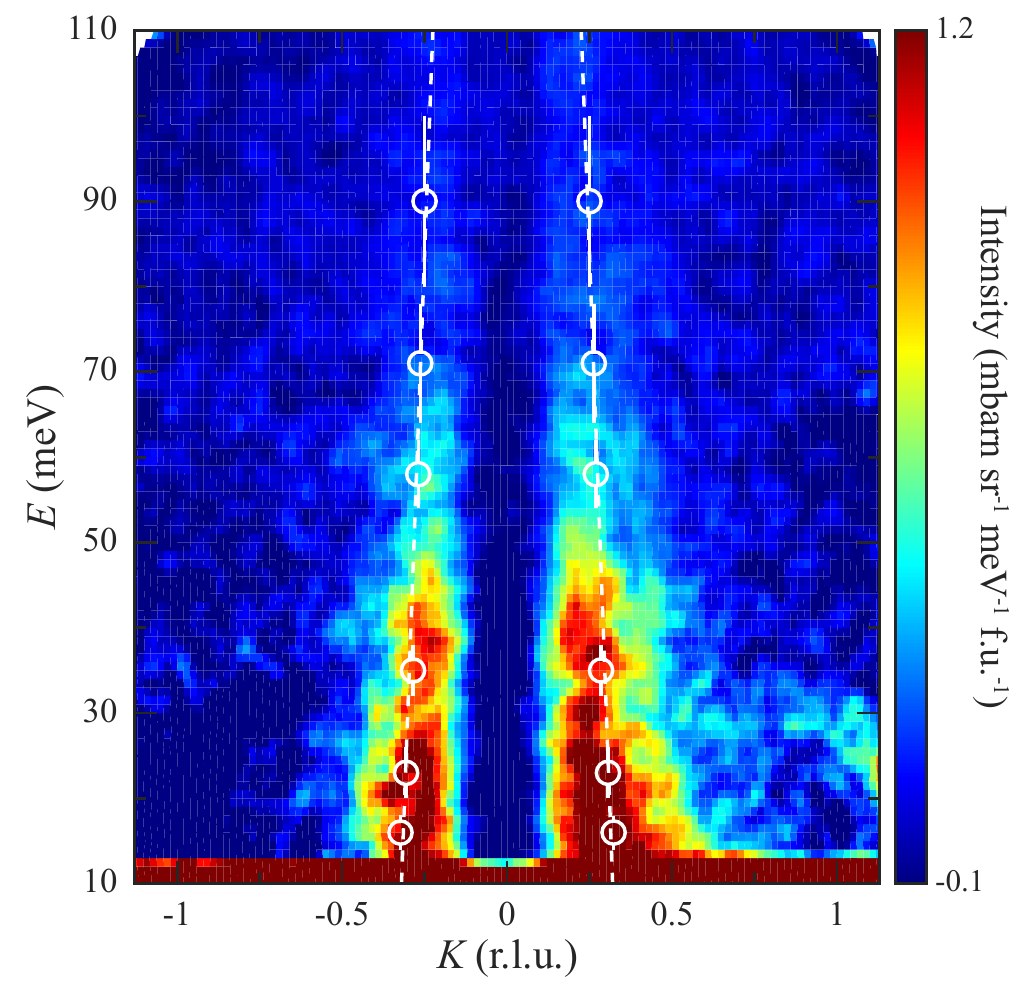}
\caption{
Dispersion of the incommensurate spin fluctuations in Sr$_{1.8}$La$_{0.2}$RuO$_4$ at 5~K measured with $E_i$ = 135.8~meV. Data are two-fold symmetrized with respect to the $K$ axis and $|{\mathbf{Q}}|$-dependent backgrounds have been subtracted. 
The intensity is integrated from $H=0.1$ to $0.5$. Open circles denote peak positions extracted from Gaussian fits of constant-energy scans in Figs.~\ref{fig4}(d)--\ref{fig4}(i). Vertical bars indicate the energy integration range and the white dashed lines are linear fits to the peak positions.
}
\label{fig3}
\end{figure}

\begin{figure*}
\centering
\includegraphics[width=0.8\textwidth]{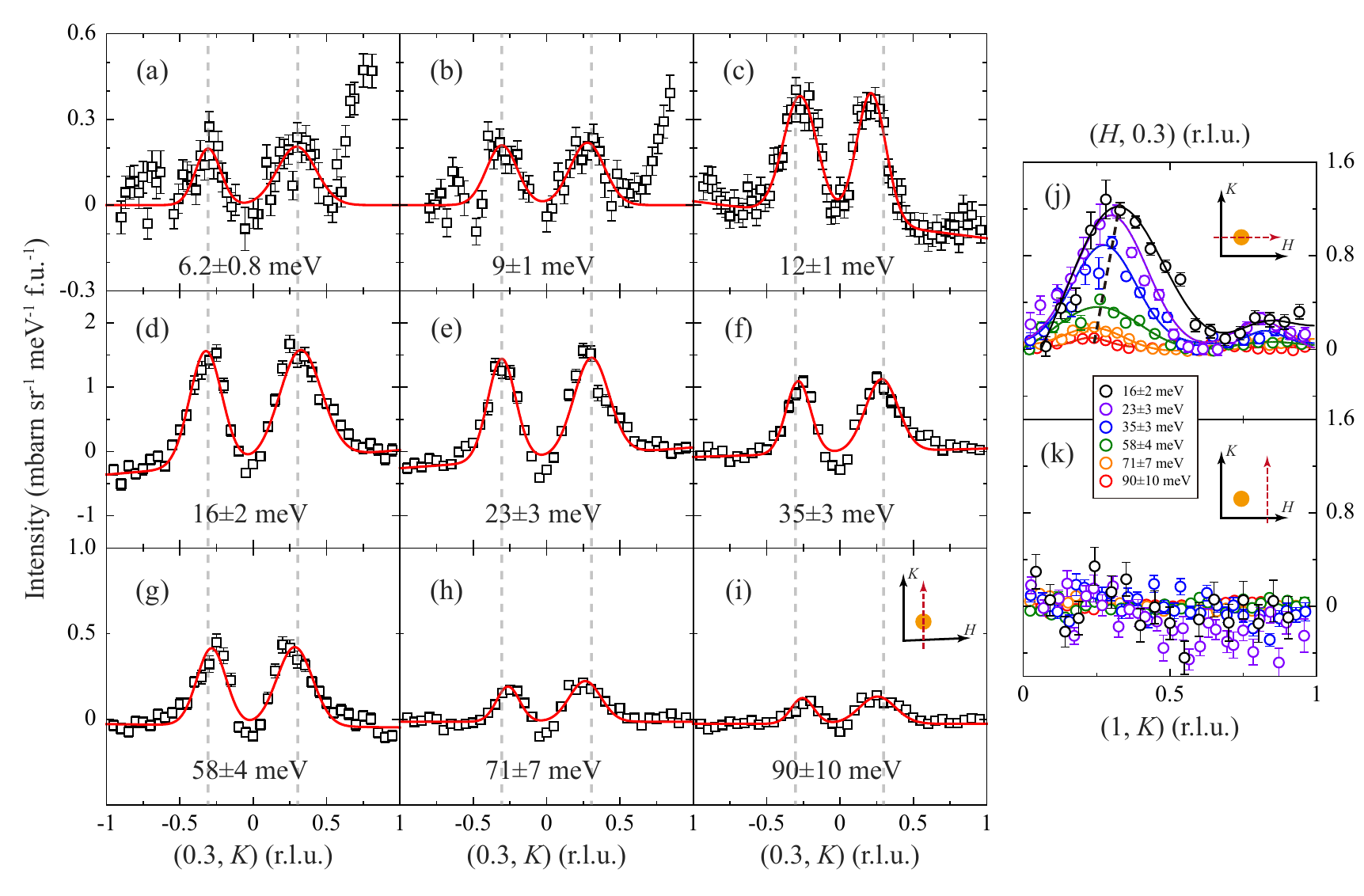}
\caption{
Constant-energy scans of the spin fluctuations in Sr$_{1.8}$La$_{0.2}$RuO$_4$ at 5~K. (a) and (b), (c), and (d)--(k) were collected on MERLIN with $E_i$ = 18.8, 40.2, and 135.8~meV, respectively. (a)–-(i) Constant-energy scans along $(0.3,K)$ direction at indicated energies. $H$ is integrated from 0.05 to 0.55. Red solid curves are fits with two Gaussian profiles and a linear background. The fitted peak positions are plotted in Fig.~\ref{fig3}. Gray dashed lines indicates $K=\pm 0.3$.
(j) Constant-energy scans along $(H,0.3)$ direction with $K$ integrated from 0.10 to 0.50. Solid lines are fits with Gaussian profiles. (k) Constant-energy scans along $(1,K)$ direction with $H$ integrated from 0.95 to 1.05. Data in (j) and (k) are symmetrized with respect to the $K$, $H$ axes and the $(1,1)$ direction to enhance statistics. Backgrounds have been subtracted as described in Ref.~\cite{WangNC2016}. The black dashed line connects the fitted peak positions at 16 and 90 meV.
Arrows in the the insets indicate the scan directions. Error bars represent one standard deviation.
}
\label{fig4}
\end{figure*}

However, some experimental data such as the absence of edge currents~\cite{KirtleyPRB2007,HicksPRB2010}, the Pauli-limited upper critical field~\cite{KittakaPRB2009}, and the lack of a linear strain dependency of $T_c$ at zero strain limit~\cite{HicksSci2014} cannot be readily explained in this scenario.
Furthermore, recent NMR~\cite{PustogowNat2019,IshidaJPSJ2020} and polarized neutron diffraction~\cite{PetschPRL2020} investigations revealed that the spin susceptibility in Sr$_2$RuO$_4$ decreases significantly below $T_c$---contradicting a simple spin-triplet Cooper pairing. 
The discrepancies between these results have kicked off a flurry of studies to identify its pairing symmetry, and different composite order parameters have been put forward to reconcile the controversies~\cite{BenhabibNP2021,GhoshNP2021,GrinenkoNP2021,GrinenkoNC2021,KivelsonNPJQM2020,SharmaPNAS2020,RomerPRL2019,RoisingPRR2019,SuhPRR2020}. Yet, the debate on this topic remains far from closed.

Superconductivity in Sr$_2$RuO$_4$ arises near magnetic instabilities~\cite{BradenPRL2002,CarloNM2012,OrtmannSciRep2013}, as it does in cuprates and iron pnictides.
As a result, spin fluctuations are believed to be critical for the Cooper pairing~\cite{MazinPRL1997,MazinPRL1999,MonthouxPRB1999,KuwabaraPRL2000,HuoPRL2013}. Theoretically, antiferromagnetic and ferromagnetic fluctuations could cause even and odd parity of superconducting order parameters, respectively~\cite{MazinPRL1999,MonthouxPRB1999}.
Previous inelastic neutron scattering (INS) experiments on Sr$_2$RuO$_4$ revealed strong two-dimensional incommensurate antiferromagnetic responses around wave vector $\mathbf{Q}_{ic}$ = (0.3,0.3,$L$) and equivalent positions~\cite{SidisPRL1999,BradenPRB2002,BradenPRL2004,IidaPRB2011,KunkemollerPRL2017,SteffensPRL2019,IidaJPSJ2020,JenniPRB2021}, which are attributed to the Fermi surface nesting between the quasi-one-dimensional $\alpha$ and $\beta$ sheets associated with the $d_{xz}$ and $d_{yz}$ orbitals, respectively [see Fig.~\ref{fig1}(a)]. 
Meanwhile, weak magnetic fluctuations appear near the Brillouin zone center~\cite{BradenPRL2004,SteffensPRL2019,JenniPRB2021}, manifesting a ferromagnetic character. The ferromagnetic fluctuations have been attributed to the nesting of $\gamma$ bands \cite{BergemannAIP2003,KikugawaPRB2004_transport}, but recent dynamical mean-field theory (DMFT) calculations instead suggested that the ferromagnetic fluctuations are quasi-local due to Hund's coupling~\cite{StrandPRB2019}. Yet, the origin of the ferromagnetic fluctuations is still under debate.

Band structure studies of Sr$_2$RuO$_4$ reveal a van Hove singularity (vHS) in the $\gamma$ band closely (${\sim}49$~meV) above the Fermi level~\cite{BergemannAIP2003,KikugawaPRB2004_dHva,KikugawaPRB2004_transport,ShenPRL2007}. Modest external perturbations can therefore significantly influence the Fermi surface topology and electronic properties~\cite{KikugawaPRB2004_dHva,KikugawaPRB2004_transport,HicksSci2014,BurganovPRL2016,SteppkeSci2017,BarberPRL2018,MarquesAdvMat2021}.
For example, applying uniaxial strain on Sr$_2$RuO$_4$ may induce a Lifshitz transition by pushing the vHS to the Fermi level and enhance $T_c$~\cite{SteppkeSci2017,BarberPRL2018}.
Carrier doping by chemical substitution provides an alternative path to tune the electronic properties.
For example, La$^{3+}$ substitution for Sr$^{2+}$ induces electron doping and generates a Lifshitz transition at the critical doping in Sr$_{2-y}$La$_y$RuO$_4$ ($y_c$ = $0.2$), where the vHS crosses the Fermi level~\cite{KikugawaPRB2004_transport,KikugawaPRB2004_dHva,ShenPRL2007}.
A non-Fermi-liquid (NFL) behavior has been observed at the Lifshitz transition, manifesting the occurrence of a strong electronic renormalization and quasiparticle scatterings [see Fig.~\ref{fig1}(b) and Ref.~\onlinecite{KikugawaPRB2004_transport}].
In the weak coupling scenario, the divergence of density of states near the vHS would greatly enhance the ferromagnetic spin fluctuations associated with the $\gamma$ band~\cite{KikugawaPRB2004_transport}, whereas quasi-local fluctuations are expected to be marginally affected and remain very weak~\cite{StrandPRB2019}.
The origin of ferromagnetic fluctuations in Sr$_2$RuO$_4$ and related materials can thus be directly addressed by studying the La$^{3+}$ doping effect. 

In this paper, we report an INS study of spin fluctuations in Sr$_{2-y}$La$_y$RuO$_4$ at the critical concentration $y_c$ = $0.2$, where the vHS in the $\gamma$ band crosses the Fermi level.  Strong incommensurate antiferromagnetic fluctuations are observed up to ${\sim}110$~meV and disperse from $\mathbf{Q}_{ic}$ = (0.3,0.3) to (0.25,0.25). This differs from undoped Sr$_2$RuO$_4$ where the antiferromagnetic fluctuations show little dispersion.
On the other hand, no discernible ferromagnetic response has been detected with our experimental sensitivity. These results contradict the weak-coupling scenario that ferromagnetic fluctuations are induced by the nesting of the $\gamma$ band; instead they support the proposal that magnetic responses near the ferromagnetic wave vectors have a quasi-local character.

High-quality Sr$_{1.8}$La$_{0.2}$RuO$_4$ single crystals were synthesized using the floating zone method~\cite{Kim2011}. Electron probe micro analysis (EPMA) and x-ray diffraction (XRD) measurements are performed to confirm the chemical composition and characterize the quality of our sample. The XRD Rietveld refinement shows that Sr$_{1.8}$La$_{0.2}$RuO$_4$ adopts the same space group ($I$4/$mmm$) as Sr$_2$RuO$_4$. No impurity, disorder, or lattice distortion is detected (see the Supplemental Material \cite{SM}).
Below 40~K, the in-plane resistivity 
follows a temperature dependence $\rho_{ab}$ = $AT^n$ + $\rho_{0}$ with $n$ = $1.54 \pm 0.04$ [Fig.~\ref{fig1}(b)], deviating from the Fermi liquid behavior observed in undoped Sr$_2$RuO$_4$. This is consistent with previous reports ~\cite{KikugawaPRB2004_transport}. Our INS experiments were carried out on the MERLIN time-of-flight spectrometer at the Rutherford Appleton Laboratory~\cite{Bewley2006}, and the cold triple-axis spectrometer PANDA at the Heinz Maier-Leibnitz Zentrum (FRMII)~\cite{Schneidewind2015}. 21 pieces of single crystals with a total mass of ${\sim}12$~g were coaligned for the INS measurements.
The time-of-flight experiment was performed with the incident neutron energies fixed at $E_i$ = 18.8, 40.2, and 135.8~meV. Data were normalized to absolute units using the incoherent elastic scattering from a standard vanadium sample. Triple-axis experiments were performed with the final neutron energy fixed at $E_f$ = 5.1 meV. Pyrolytic graphite $(002)$ [PG$(002)$] was used as a monochromator and analyzer. A Be filter was employed to reduce the contamination from higher-order neutrons. We define the wave vector $\bf{Q}$ in the tetragonal unit cell at $\bf{Q}$ = $H\bf{a^*}$+$K\bf{b^*}$+$L\bf{c^*}$ as ($H$, $K$, $L$) in reciprocal lattice units (r.l.u.), where $\bf{a^*}$ = $2\pi$\bf{\^{a}}$/a$, $\bf{b^*}$ = $2\pi$\bf{\^{b}}$/b$ and $\bf{c^*}$ = $2\pi$\bf{\^{c}}$/c$ \rm with lattice parameters $a$ = $b$ = 3.86 \AA\ and $c$ = 12.72 \AA.

Figure~\ref{fig2} displays constant-energy plots of the scattering intensity in the $(H,K)$ plane at 5~K. At low energies [Figs.~\ref{fig2}(a)--\ref{fig2}(c)], clear scatterings are observed near the incommensurate wave vector $\mathbf{Q}_{ic}$ = (0.3,0.3) and equivalent positions. This is similar to the dominant antiferromagnetic spin fluctuations in Sr$_2$RuO$_4$~\cite{SidisPRL1999,BradenPRB2002,IidaPRB2011}. 
It has been shown that La doping mostly influences the $\gamma$ band, while the nesting condition between the $\alpha$ and $\beta$ Fermi sheets is weakly affected~\cite{ShenPRL2007}.
The scattering significantly weakens at the equivalent $\mathbf{Q}=(0.7,0.3)$ [Figs.~\ref{fig2} and \ref{fig4}(j)] due to the reduced {Ru}$^{4+}$ magnetic form factor with increasing $|\mathbf{Q}|$, corroborating its magnetic origin.
With increasing energy, the incommensurate peak positions move toward the lower $\mathbf{Q}$ [Figs.~\ref{fig2}(d)--\ref{fig2}(i)], revealing a notable dispersion.

The dispersion of the incommensurate spin fluctuations can be better visualized in the contour plot of the $E$-$K$ plane. As illustrated in Fig.~\ref{fig3}, strong spin fluctuations stem from $\mathbf{Q}_{ic}$ = (0.3,0.3) and disperse to $(0.25,0.25)$ around 110~meV.
This behavior differs from Sr$_2$RuO$_4$ where the peak position of spin fluctuations is essentially fixed at $\mathbf{Q}_{ic}$ at all energies~\cite{IidaPRB2011}. This difference indicates that La doping may induce a non-negligible change to the low-energy dispersion of the $\alpha$ and $\beta$ bands that slightly departs from the rigid-band shift assumption~\cite{KikugawaPRB2004_transport}. 
Note that since the energy transfer is coupled to $L$ in the time-of-flight scattering geometry, the $E$-$K$ contour plot yields no observable $L$ modulation of the spin fluctuations.

We made constant-energy cuts through $\mathbf{Q}$ = (0.3,0.3) and (1,0) to quantify the spin fluctuations.
As shown in Figs.~\ref{fig4}(a)--\ref{fig4}(j), the intensity of antiferromagnetic fluctuations displays a maximum at around 16~meV and gradually vanishes above 110~meV. The peak position shifts to a lower $\mathbf{Q}$ with increasing energy in a nearly linear fashion [Figs.~\ref{fig3} and \ref{fig4}(j)].
A careful survey of the scattering intensity has been done around ferromagnetic wave vectors $(1,0)$ and $(1,1)$. However, scans in the $(1,K)$ direction covering these wave vectors are featureless at all energies [Fig.~\ref{fig4}(k)], implying that ferromagnetic fluctuations, if they exist, are still below the detection limit of the current INS instrument. This is further confirmed by our  triple-axis measurements (see Fig. S3. in the Supplemental Material \cite{SM}).

Previous polarized neutron scattering measurements revealed relatively weak ferromagnetic fluctuations in Sr$_2$RuO$_4$~\cite{BradenPRL2004,SteffensPRL2019,JenniPRB2021}. 
These studies, however, were not able to distinguish whether the ferromagnetic fluctuations originate from itinerant electrons or local moments.
Tight-binding calculations predict a divergence of bare band susceptibility $\chi_0$ and an even more pronounced enhancement of the renormalized susceptibility $\chi$, when La doping pushes the vHS to the Fermi level and induces a divergence of the density of states in the $\gamma$ band ~\cite{BergemannAIP2003,KikugawaPRB2004_transport}.
Indications of such enhanced electronic density of states were indeed observed by transport, specific heat and de Haas-van Alphen oscillations measurements~\cite{KikugawaPRB2004_transport,KikugawaPRB2004_dHva}.
Our data, however, reveal no discernible signal around the ferromagnetic wave vectors $\mathbf{Q}=(1,0)$ and $(1,1)$ at all energies measured, contradicting this weak-coupling scenario.
Alternatively, the magnetic response around $\Gamma$ point with a broad momentum distribution has been interpreted as quasi-local fluctuations driven by Hund's coupling in the DMFT calculation~\cite{StrandPRB2019}. The absence of a sharp enhancement of the ferromagnetic fluctuations in Sr$_{1.8}$La$_{0.2}$RuO$_4$ seems in favor of such a proposal. Interestingly, a recent angle-resolved photoemission measurement on the monolayer SrRuO$_3$ film also suggested that the vHS in the $\gamma$ band does not lead to ferromagnetism but results in a strongly correlated metal state \cite{SohnNC2021}. This is in line with current neutron scattering results in bulk Sr$_{1.8}$La$_{0.2}$RuO$_4$.

The prevalent paradigm of a $p$-wave triplet pairing state involves the exchange of bosonic ferromagnetic modes, while antiferromagnetic fluctuations favor a singlet pairing. The lack of conventional dispersive ferromagnetic fluctuations in Sr$_2$RuO$_4$ seems to make it less likely to host a simple $p$-wave pairing than other candidates, where substantial well-defined ferromagnetic fluctuations have been observed, such as the heavy Fermion superconductor UCoGe~\cite{StockPRL2011} and iron-based compound YFe$_2$Ge$_2$~\cite{WoPRL2019}. Recent theoretical and experimental works have suggested composite order parameters in Sr$_2$RuO$_4$ \cite{BenhabibNP2021,GhoshNP2021,GrinenkoNP2021,GrinenkoNC2021,KivelsonNPJQM2020,RomerPRL2019}, which may involve multiple spin fluctuation pairing channels.  
Sophisticated theoretical calculations considering both antiferromagnetic and quasi-local ferromagnetic fluctuations on the Cooper pairing are highly desirable.

In summary, we used INS experiments to investigate spin fluctuations in Sr$_{1.8}$La$_{0.2}$RuO$_4$ covering a broad range of energy-momentum space. Strong antiferromagnetic spin fluctuations emanate from the incommensurate position $(0.3,0.3)$. The excitation extends to above 110~meV and progressively shifts to $(0.25,0.25)$.
On the other hand, there is no clearly increased scattering near ferromagnetic wave vectors.  
This contradicts the weak-coupling hypothesis, which predicts a divergence of ferromagnetic susceptibility as the vHS approaches Fermi level at this La doping.
Instead, the inertia of weak ferromagnetic fluctuations against La doping is in line with the description of quasi-local fluctuations driven by Hund's coupling~\cite{StrandPRB2019}.
These results are crucial for a complete understanding of the magnetism in this system, and they set significant restrictions  on the mechanism of superconductivity in Sr$_2$RuO$_4$.\\

Acknowledgements:
This work was supported by the Key Program of the National Natural Science Foundation of China (Grant No. 12234006), the National Key R\&D Program of China (Grant No. 2022YFA1403202), and the Shanghai Municipal Science and Technology Major Project (Grant No. 2019SHZDZX01). Q.W. was supported by the CUHK Research Startup Fund (Grant No. 4937150). Y.F. was supported by Xie Jialin Youth Foundation of Institute of High Energy Physics (Grand No.
E2546JU2), and National Key R\&D Program of China (Grand No. 2022YFA1604104). H.W. acknowledges support from the China National Postdoctoral Program for Innovative Talents (Grant No. BX2021080), China Postdoctoral Science Foundation (Grant No. 2021M700860), the Youth Foundation of the National Natural Science Foundation of China (Grant No. 12204108), and Shanghai Post-doctoral Excellence Program (Grant No. 2021481). The work at SNU was supported by the Institute for Basic Science in Korea (Grant No. IBS-R009-G2). D.T.A. thanks EPSRC-UK (Grant No.EP/W00562X/1). Experiments at PANDA were conducted under proposal numbers 5114 and 5719. Datasets collected at MERLIN spectrometer are available from the ISIS facility, Rutherford Appleton Laboratory data portal (https://doi.org/10.5286/ISIS.E.42581971).

\end{document}